\documentclass[aps,prl,twocolumn,amsmath,amssym,amsthm,superscriptaddress,reprint]{revtex4-1}
\pdfoutput=1
\usepackage{graphicx}
\usepackage{dcolumn}
\usepackage{bm}
\usepackage{url}
\usepackage{bm}   
\usepackage{bbm}   
\usepackage{verbatim}
\usepackage{stmaryrd}
\usepackage{amsthm}
\usepackage{amssymb}
\usepackage{array,multirow}
\usepackage{empheq}
\usepackage[usenames,dvipsnames]{color}
\usepackage[normalem]{ulem}
\usepackage{tikz}

\theoremstyle{plain}

\theoremstyle{definition}

\theoremstyle{remark}

\newcommand{\LS}{Szil\'{a}rd}

\begin{document}
\title{Thermodynamic cost and benefit of memory}
\author{Susanne Still}
\affiliation{Department of Information and Computer Sciences, and Department of Physics and Astronomy, University of Hawaii at M\=anoa, sstill@hawaii.edu}
\begin{abstract}
This letter exposes a tight connection between the thermodynamic efficiency of information processing and predictive inference. 
A generalized lower bound on dissipation is derived for {\em partially observable} information engines which are allowed to use temperature differences. 
It is shown that the retention of {\em irrelevant} information limits efficiency. {
A data representation strategy is derived from optimizing a fundamental physical limit to information processing: minimizing the lower bound on dissipation leads to a data compression method that maximally retains relevant, predictive, information. In that sense, predictive inference emerges as the strategy that least precludes energy efficiency}. 

\end{abstract}
\maketitle
Effective and meaningful processing of information is crucial for the operation of 
``Maxwell's demon", a thought experiment in which a sentient being operates a trap door in a split gas container to sort fast from slow molecules, attempting to defy the Second Law of thermodynamics \cite{maxwell-demon}. The thermodynamic cost of the demon's operation offsets any gains \cite{brillouin1951maxwell}, ensuring that the Second Law is not violated. 
Details of this process have been discussed for over a hundred years \cite{demonbook90}, but recently, interest in understanding the thermodynamics of information processing has spiked with the increasing ability to control bio-molecular machines, and the advent of nanotechnology \cite{sagawa2009minimal, berut2012expLandauer, mandal2012work, SagawaUedaFeedback12, mandal2014, koski2014experimental, exp-landauer2014, parrondo2015thermodynamics, martinez2016brownian, hong2016experimental, gavrilov2016erasure, gavrilov2017direct, lathouwers2017memory, kumar2018nanoscale, paneru2018lossless, admon2018experimental,wolpertbook2019}.

Ninety years ago, a simple and elegant thought experiment was proposed by L{\'e}o \LS\ \cite{szilard} to illuminate the concept behind Maxwell's demon without having to discuss any detailed physiology of a sentient being. \LS\ considered a one-particle gas in a container with a movable wall in the middle. Knowing which side is empty allows for work extraction by isothermal expansion. Highly influential to this day (e.g. \cite{zurek1986maxwell, magnasco1996szilard, lloyd1997quantum, kim2011quantum, vaikuntanathan2011modeling, proesmans2015, brittain2019biochemical}), \LS's simple information engine
forms a basis for understanding the conversion between information and work, and fundamental thermodynamic limits to computation.

Importantly, only information about certain aspects of a system at hand can enable work extraction. What these aspects are is determined by the physics of any given setup. An observer has to make choices about what to measure, what part of the observable data to store in memory, and to which precision. The knowledge so acquired is then used to act. Interactive observers are often called ``agents" in machine learning and robotics. 
\LS\ assumed all agent choices to be optimal, and all relevant degrees of freedom to be observable. These assumptions reverberate in the ensuing literature, limiting the discussion largely to the special case in which {\em all} of the captured information can, in principle, be turned into work
\footnote{It has been discussed that measurement errors limit how much information can be obtained, and thus used, e.g. in \cite{SagawaUedaFeedback12},  but here we focus on structural constraints {that dictate limited exploitable correlations between the observable data and} the variables that need to be known to extract work.}.

But in many real-world scenarios encountered by agents, biological and artificial alike, not all degrees of freedom of an observed system are accessible, thus fundamentally limiting agents to act on partial knowledge. In the most general case, systems are only partially observable{, and the agent has to make inferences to predict relevant quantities from available data. This predictive inference constitutes a core function of intelligent behavior. The relative costs and benefits of more or less complicated memories have to be weight against each other.}
Intuitively, smarter choices about how to represent observations in memory should result in fewer losses, but how precisely should a general strategy be designed? Can it be derived from a physical principle? 

This paper reveals that {performing} predictive inference {by discarding irrelevant information} is a strategy that not only enables {energy efficiency}, but also emerges naturally from minimizing a lower bound on dissipation---demonstrating that an intelligent data representation strategy can be derived from the optimization of a fundamental physical limit to information processing. 

\paragraph{Information Engines.}
A variety of models have been studied which use information to extract work from a heat bath. This information either is acquired and memorized \cite{szilard}, or it is supplied from the outside, whereby the costs of running a memory can be ignored, and bits can be viewed as fuel \cite{mandal2014}. We follow \LS's lead, including the cost of information acquisition and decision-making in the engine's energy bill. We generalize to partial observability and allow the use of temperature differences. A lower bound on dissipation encountered by generalized information engines is given (Eq. \ref{result}), together with a \LS-Carnot cycle that saturates the bound. 

An information engine contains: 1) a (partially) observable system. The system's microstate shall be denoted by the random variable $Z$ with realizations $z \in {\cal Z}$. 2) An agent, implemented with another physical system (artificial or biological) that turns measurements into a stable memory, denoted by the random variable $M$ with realizations $m \in {\cal M}$. The memory is used to decide on a work extraction protocol. 3) A work extraction device that enables the agent to couple useful energy out of the system. This device is given by the physical setup, and determines which aspects of the system are relevant with respect to extracting work. 

An information engine runs cyclicly. A measurement is performed at time $t^i_0$, and written into memory ($i$ denoting the $i$-th cycle; depending on the context, we may later drop the superscripts). This process 
is implemented by a protocol 
that changes external control parameters 
on the memory between $t_0^i$ and $t_1^i$. The protocol is a function of the observable data at time $t_0^i$.
During this process, the engine is connected to a heat bath at temperature $T$. An average amount of work $W_M$ is done on the memory, and an average amount of heat, $-Q_M$, is dissipated (the convention is used that energy flows {\em into} a system are positive). The memory 
remains unchanged until $t^{i+1}_0$. 

Work extraction starts at time $t^i_2$, and ends at $t^i_3$: a protocol is chosen as a function of the memory state and applied to the system.
As a result, heat, in the average amount of $Q_E$, is absorbed from a heat bath at temperature $T'$, whereby an average amount of work, $-W_E$, is extracted. This is implemented using whichever work extraction device is given---in \LS's example it is the movable partition, combined with pulleys and weights \cite{szilard}. The work extraction protocol is designed such that the memory has no exploitable correlations with the system after $t^i_3$. 

The system is then prepared for the next cycle between times $t_3^i$ and $t_0^{i+1}$. This preparation step is always the same procedure, and hence
independent of the memory state---in \LS's example the partition gets re-inserted into the middle of the box. Typically, it is assumed not to require work. 
{At the start of the new cycle, the old memory state contains no correlations with the new state of the system. The memory is not reset to a specific state.}

To calculate the engine's free energy changes during steps of this cycle, we consider
the generalized free energy \cite{shaw1984dripping, GavSchul97, Crooks07, takara2010generalization, deffner2016quantum, lostaglio2015description},
$F_{t} = \langle E_t(m,z)\rangle_{p_t(m,z)} - k T H_t$,
where $p_{t}(m,z)$ denotes the joint distribution, over system states, $z$, and memory states, $m$, at time $t$, $E_t(m,z)$ the energy, and $H_t = - \left\langle \ln\left[{p_t(m,z)}\right] \right\rangle_{p_t(m,z)}$ the Shannon entropy. (The shorthand $p(x)\equiv p(X\!=\!x)$ is used, and $\langle \cdot \rangle_{p(x)}$ is the average over $p(x)$.) Changes occur due to manipulation of the memory, $\Delta F_{M} \equiv F_{t_1} - F_{t_0}$, and due to work extraction, $\Delta F_{E} \equiv F_{t_3} - F_{t_2}$. The associated free energy differences, 
\begin{eqnarray} 
&& \Delta F_{A}  = W_A + Q_A - kT\Delta H_{A}~; A\in \{M,E\} ~, \label{DF}
\end{eqnarray}
contain an average change in energy, $\Delta E_{A} = W_A + Q_A$, which is the sum of the average work and heat, according to the First Law of thermodynamics, and they contain entropy changes, $\Delta H_{M} \equiv H_{t_1} - H_{t_0}$ and $\Delta H_{E} \equiv H_{t_3}- H_{t_2}$.
To calculate those, it is useful to introduce the following notation:
\paragraph{\bf 1. Decomposition:} We will decompose the microstate of the system in two different ways, first, into observable components, $X$, versus non-observable components, $\bar{X}$: $Z=(X,\bar{X})$ \footnote{Random variables are denoted by capital letters, realizations by small letters.}. 
However, what can be observed is not necessarily the same as what can be manipulated. Denote by $Y$ all components with the properties: 1) they can be manipulated between $t_2$ and $t_3$, and 2) they are predictable from $t_0$ to $t_2$ (the mutual information is nonzero). These are the components relevant for work extraction. $Z$ can also be decomposed into those, versus all other components, $\bar{Y}$: $Z=(Y,\bar{Y})$.

\paragraph{\bf 2. Manipulation:} The system cannot be manipulated in a way that changes anything but $Y$. Thus, 
\begin{equation}
p_{t_2}(\bar{y}|y,m) = p_{t_3}(\bar{y}|y,m) \equiv p(\bar{y}|y,m). \label{manip}
\end{equation}

\paragraph{\bf 3. Data representation:} The stochastic map from observable data, $x$, to memory states, $m$, is independent of $\bar{x}$: 
\begin{equation}
p_{t_1}(m|z) = p_{t_1}(m|x,\bar{x}) = p_{t_1}(m|x) \equiv p(m|x). \label{dr}
\end{equation}

\paragraph{\bf 4. Marginal distributions:} 
All changes performed on the system 
do not change the marginal distribution 
at times $t_{k}$ ($k = 1, \dots, 4$): $p_{t_{k}}(z) \equiv p(z)$, and therefore, $p_{t_k}(y) = \sum_{\bar{y}} p_{t_k}(y, \bar{y}) \equiv p(y)$ 
and $p_{t_k}(x) \equiv p(x)$.
The preparation step simply introduces a hidden variable $v$, such that $p_{t_0}(y) = \sum_v p(y|v)p(v)$. 
If $v$ is discovered, then the system appears in a non-equlilibrium state best described by $p(y|v)$,
which can be exploited during work extraction. 

The marginal probability of the memory derives only from the statistical average of measurement outcomes. Therefore,  
$p_{t_k}(m) = \sum_x p(m|x) p(x) \equiv p(m). \label{pM}$

\paragraph{\bf 5. Inference:} Knowledge carried in memory about the relevant quantity $Y$ derives from the statistical average over measurement outcomes: 
\begin{eqnarray}
&p_{t_2}(y|m) = \sum_x p(y|x)p(m|x)p(x) \equiv p(y|m)  ~. \label{inference}
\end{eqnarray}
Here we used that if the measurement outcome itself is given, then the memory adds no information about the relevant quantity ($p(y|m,x)\!=\!p(y|x)$). 
The dynamics of the system determine $p(y|x)$.

\paragraph{Thermodynamic cost of memory.}
At $t_0$, system and memory are uncorrelated and the joint distribution factorizes: 
$p_{t_0}(m,z)\!=\! p_{t_0}(m)p_{t_0}(z)\! =\! p(m)p(z)$.
After the memory is constructed, 
$p_{t_1}(m,z) = p_{t_1}(m|x,\bar{x})  p_{t_1}(z) = p(m|x)p(z)$.
The entropy thus decreases by $\Delta H_M = H[M|X] - H[M] = -I[M,X]$, the amount of mutual information captured in the memory about the observable data. This process happens at temperature $T$ and the free energy change associated with it, Eq. (\ref{DF}), is 
$\Delta F_M = W_M + Q_M +kT I[M,X] ~.$
The Second Law implies that $W_M - \Delta F_M \geq 0$,
and hence 
\begin{equation}
- Q_M \geq kT I[M,X]~. \label{boundQm}
\end{equation}
Operating a memory requires at a minimum dissipation proportional to the amount of information retained. This is a known fact (e.g. \cite{Landauer1961, parrondo2015thermodynamics, Still-LiL} and references therein).  

\paragraph{Thermodynamic gain derivable from a memory.}  
At the beginning of work extraction, the joint distribution, written with the decomposition $Z=(Y,\bar{Y})$, is
\begin{eqnarray}
&p_{t_2}(m,z) &=\!  p_{t_2}(\bar{y}|y,m) p_{t_2}(y|m) p_{t_2}(m)\! =\! p(\bar{y}|y, m) p(y|m) p(m). \notag
\end{eqnarray}
At the end of work extraction, all correlations between $Y$ and $M$ are gone, and we have
\begin{eqnarray}
&p_{t_3}(m,z) &=\! p_{t_3}(\bar{y}|y,m) p_{t_3}(y) p_{t_3}(m)\! =\! p(\bar{y}|y, m) p(y) p(m). \notag
\end{eqnarray}
The entropy of the joint system thus increases by $\Delta H_E = H[Y] - H[Y|M] = I[M,Y]$.
During this process, the engine is connected to a bath at temperature $T'$. The free energy change is
$\Delta F_E = W_E + Q_E -kT' I[M,Y]$, 
and the second law implies that $W_E - \Delta F_E \geq 0$.
Therefore,  
\begin{equation}
Q_E \leq kT' I[M,Y] ~. \label{boundQe}
\end{equation}
The amount of heat that can get absorbed and turned into work is limited by how much information is retained about the variable(s) relevant with respect to work extraction. If the relevant quantities are fully observable, then all of the energetic cost of running the memory can in principle be recovered.
But, in general, {the information captured in memory, \mbox{$I_{\rm mem} \!\equiv\! I[M,X]$}, may contain only some bits which are predictive of the relevant quantity, {that is,} $I_{\rm rel}\! \equiv\! I[M,Y]$, relevant, or predictive, information.} {The rest is} {\em irrelevant information}, \mbox{$I_{\rm irrel}\! \equiv\! I_{\rm mem}\! -\! I_{\rm rel} \geq 0$}. 
{This quantity is non-negative because $p(y|m,x)=p(y|x)$ implies that $I[M,X] - I[M,Y] = I[M,X|Y] \geq 0$. We thus see that $Q_E \leq kT' I[M,Y] \leq kT' I[M,X]$, i.e., inequality (\ref{boundQe}) is more restrictive than the known ultimate bound, $Q_E \leq kT' I[M,X]$ (e.g. \cite{sagawa2008second,parrondo2015thermodynamics,strasberg2017quantum} and references therein), which can be reached only in the special case where the relevant quantities are fully observable.} 
\paragraph{Lower bound on dissipation.}
\noindent
Over a cycle, an isothermal information engine (connected to a heat bath at temperature $T$ throughout) dissipates an average amount of heat $-Q\! =\! -Q_M -Q_E$.
From Eqs. (\ref{boundQm}) and (\ref{boundQe}), substituting $T'=T$, we see that the irrelevant information retained in memory sets a limit on how small dissipation can be:
\begin{equation}
- Q \geq kT \; I_{\rm irrel} \geq 0~. \label{isothermal-diss}
\end{equation}

\paragraph{Example.}
We recall \LS's Gedankenexperiment \cite{szilard} to build on it: one particle is trapped in a rectangular container with volume $V$. Let us look at the container from above, and insert a movable partition in the middle, at $y=0$, parallel to the x-axis. Knowing the sign of the particle's y-position allows us to extract work up to $kT \ln(2)$. But if we limit an agent to {\em only} observe the x-position of the particle, then no work can be extracted, because we cannot infer from the x-position alone which half of the container is empty. 

\begin{figure}[h]
\centering 
\includegraphics[width=0.95\linewidth]{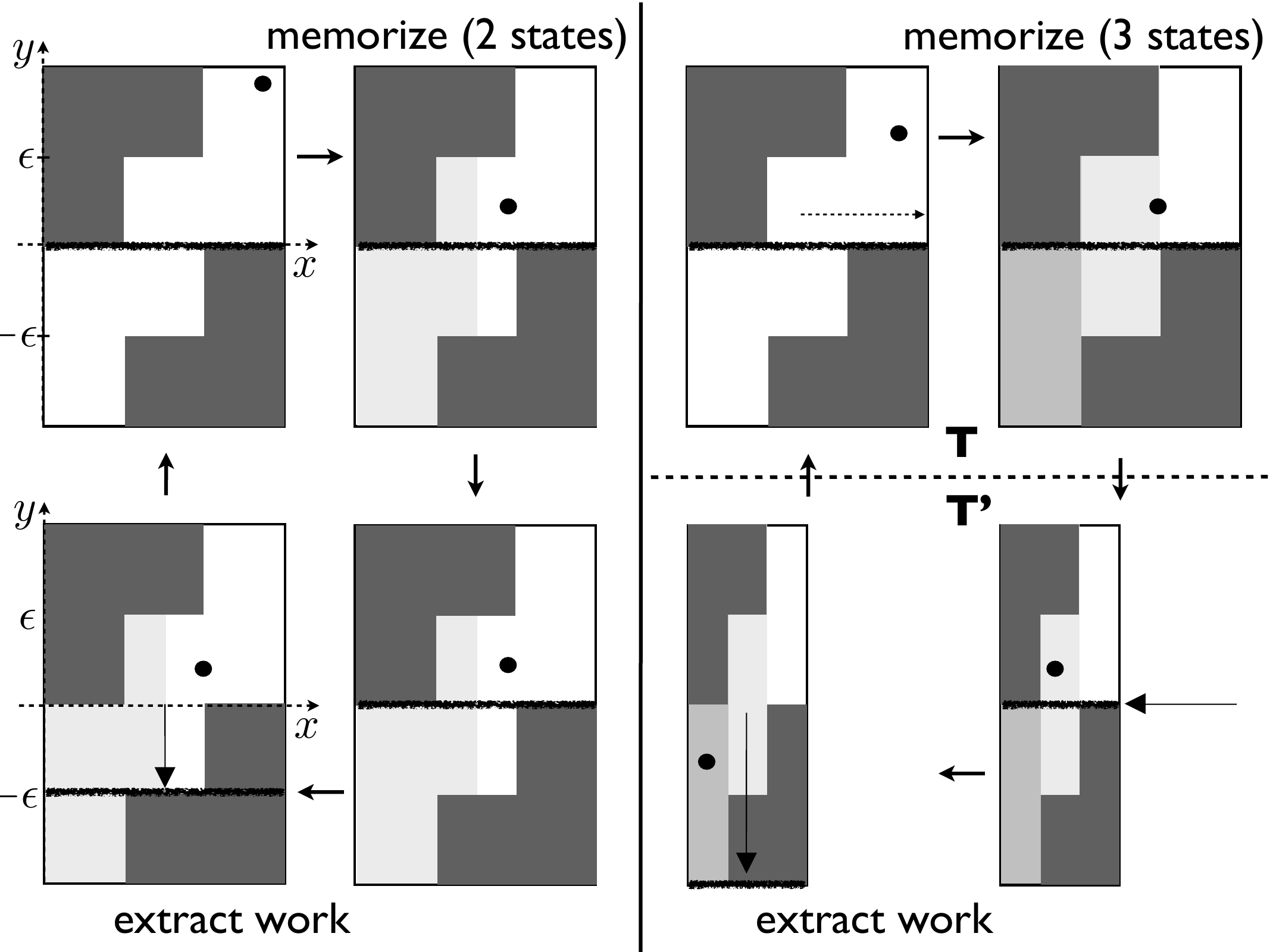}
\caption{Sketch of a partially observable \LS\ engine, run isothermally with a two-state memory (left panel), or with a three-state memory (right panel)  using a temperature difference.}
\label{stairs}
\end{figure}
Correlations can be introduced by making some areas inaccessible to the particle (Fig. \ref{stairs}, very dark grey). 
Here, the shape of the excluded areas is chosen to make it easy for the reader to calculate $I[X,Y]$ \footnote{This is perhaps a bit crude, a more natural shape might be a triangular area which would be equivalent to tilting the partition in the original \LS\ box by an angle, where the angle determines how much relevant information is accessible from knowing the y-position.}. 
The lighter grey, and white areas in Fig. \ref{stairs} signify that coarse graining along the x-axis, which corresponds to memory states. We distinguish between two (left), or three (right) different states.

To extract work, we infer if the particle is in the upper or lower half, using the memory state $m$ via the inference $p(y|m)$. The wall is then moved (reversibly) to the side that is empty with higher probability. This will result in compression, rather than expansion, with probability $q(m)$. Therefore, we have to leave a fractional volume, $\rho(m) V$, unused. Given $m$, we extract, on average, $k T' \bigl((1-q(m))\ln\left[{V -\rho(m) V \over V/2}\right] + q(m) \ln\left[{\rho(m) V\over V/2}\right]\bigr)$. This is maximized by choosing $\rho(m)\!=\! q(m)$, resulting in a total average extracted work of $-W_E=k T' \sum_m p(m) \bigl(\ln(2)+\left(1-q(m)\right)\ln\left[1-q(m)\right]+q(m) \ln\left[q(m)\right]  \bigr)=k T' (H[Y] -  H[Y|M])=kT' I[M,Y]$. Because this isothermal transformation leaves the average energy of the one-particle gas unchanged, it saturates the bound in Eq. (\ref{boundQe}), as $-W_E = Q_E = kT' I[M,Y]$.

The particle's x-position at time $t_0$, which is the observable data, $X$, contains $I[X,Y] = {2\over 3} \ln(2)$ nats of relevant information that can be fully captured by memorizing in which third along the x-axis the particle was found (right panel, upper right drawing). This costs at least $kT\ln(3)$, and allows us to extract at most ${2\over 3}kT\ln(2)$: 2/3 of the time we know for sure which side is empty (grey/white), otherwise (light grey) we can not extract any work.

If the temperature is fixed, then dissipation encountered with a three-state memory is at least $kT(\ln(3)-2\ln(2)/3)\simeq 0.64 kT$ (Eq. (\ref{isothermal-diss})). Less dissipation can be encountered with a two-state memory, costing at least $kT \ln(2)$, and capturing $I[M,Y] = {5 \over 6} \ln(5) -\ln(3)$, all of which can be converted to work if 1/6 of the total volume is left unused, e.g. by moving the wall between $\pm \epsilon$ in \mbox{Fig. \ref{stairs}}, far left panels. This results in an average dissipative loss of at least \mbox{$kT(\ln(6)-5 \ln(5)/6) \simeq 0.45 kT$}.

While the more detailed three-state memory allows us to extract more work than the less detailed two-state memory, the relative benefit does not outweigh the relative increase in cost. 

\paragraph{Access to two temperatures.}
The situation changes if we allow the engine to form a memory at temperature $T$, but extract work at a higher temperature $T'$.  
Whether it is advantageous to use a two-state or a three-state memory now depends on the temperature ratio $\alpha \!\equiv\! T'/T \!>\!1$. 
The added gain of the three-state memory outweighs the additional cost when 
$\alpha \!>\! {\ln(3) -\ln(2) \over \ln(3)+2\ln(2)/3 -5\ln(5)/6}\! \simeq\! 1.847$.

While an isothermally run information engine can, at best, recover the energy needed to run the memory, an engine that has access to two different temperatures can produce non-negative work output. The engine needs to be heated from $T$ to $T'$ between $t_1^i$ and $t_2^i$, and cooled from $T'$ to $T$ between $t_3^i$ and $t_0^{i+1}$. We consider ``memory preserving" temperature changes that do not destroy the correlations between memory and system. No additional degrees of freedom get unlocked at the higher temperature, and heating and cooling steps taken together cannot result in a net influx of heat, so that the total heat dissipated on average over a cycle is $-Q \geq -Q_M -Q_E$. The bound in Eq. (\ref{isothermal-diss}) is thus generalized (with Eqs. (\ref{boundQm}) and (\ref{boundQe})): 
\begin{equation}
- Q \geq k\left(T I_{\rm mem} - T' I_{\rm rel} \right)~. \label{result}
\end{equation}
Below, we give a protocol, akin to a Carnot process, that saturates the bound.

\paragraph{Finding a thermodynamically efficient information processing strategy.}
The bound in Eq. (\ref{result}) directly informs an optimal way to represent available data. Recall that the observer's strategy is characterized by the stochastic map $p(m|x)$. This map determines not only the amount of memory retained, but also how much relevant information is captured. We can now find the best strategy as that which {\em allows for the smallest possible dissipation}, by minimizing the bound in Eq. (\ref{result}). Mathematically, the optimization is equivalent to:  
\begin{eqnarray} \label{IB}
&&\min_{p(m|x)} \left( I[M,X] - {\alpha} I[M,Y] \right)  \\
&& {\rm subject \; to:} \;\;  \sum_m p(m|x), \; \forall x.
\end{eqnarray}
The added constraints simply ensure normalization of the stochastic map $p(m|x)$. 
This optimization problem is a lossy data compression method called the Information Bottleneck \cite{IBN}. It is a fairly general, information theoretic, predictive inference method, containing within it other methods as special cases ({for a review} see e.g. \cite{Still-IBPI-2014}).
The temperature ratio $\alpha$ determines the relative importance of the added benefits of a more detailed data representation, as we have seen in the example. It can also be interpreted as a Lagrange multiplier that controls the trade-off between a concise summary and keeping relevant information \cite{IBN}.  

\paragraph{A protocol to change between temperatures.}
A generalized \LS\ information engine can be run in a process akin to the Carnot process.
The data representation step is the same as before, an isothermal transformation at temperature $T$.
This step can be implemented reversibly if $\Delta E_M = 0$, implying that $W_M = - Q_M = kT I_{\rm mem}$.
 
A memory dependent work extraction protocol is then chosen (see right side of Fig. \ref{stairs}), consisting of the following steps: 1) isolate the one-particle gas box from the heat bath and perform an isentropic (reversible and adiabatic) compression of the entire box from volume $V$ to $V'$, raising the temperature to $T'$; 2) connect to a heat bath at $T'$, and extract work isothermally by moving into the most probably empty direction, leaving an optimized fractional volume; 3) isolate the box and perform an isentropic expansion from $V'$ to $V$, lowering the temperature back to $T$, and 4) remove the partition and re-insert in the middle. The work done in the isentropic steps one and three cancels, and overall this protocol turns the maximally possible amount of heat into work, $Q_E= -W_E= kT' I_{rel}$. 

The temperature ratio is set by the adiabatic equation, implying that $\alpha = V/V'$. 
In the right panel of \mbox{Fig. \ref{stairs}}, adiabatic compression decreases the volume by more than the factor $\simeq 1.847$, thus making the sketched three-state memory more efficient than the two-state memory. 

\paragraph{Engine Efficiency.}
This process saturates the lower bound on dissipation, Eq. (\ref{result}). It produces net work in the amount of $-W_E - W_M = k\left(T' I_{\rm rel} - T I_{\rm mem} \right)$, and absorbs heat at the higher temperature in the amount of $Q_E=kT' I_{\rm rel}$. Thus, it has an efficiency of
\begin{equation}
\eta = 1 - {T\over T'}  { I_{\rm mem} \over I_{\rm rel} } = \eta_C - {T\over T'} { I_{\rm irrel} \over I_{\rm rel}}~. \label{efficiency}
\end{equation}
The Carnot efficiency, $\eta_C = 1-{T\over T'}$, is reduced in proportion to the ratio of irrelevant to relevant information.
Efficiency in regular heat engines is non-negative. Information engines encounter a loss in the memory making process that can outweigh any thermodynamic gain of having the memory, and efficiency is non-negative only when the fraction of relevant bits retained in memory is larger than the fraction of low to high temperature: $I_{\rm rel}/I_{\rm mem} \geq T/T'$. 

{The better an information engine is at solving the predictive inference task by memorizing relevant bits and eliminating irrelevant bits, the more efficient it can be}. 

\paragraph{Discussion.}
We have seen that it is possible to derive a rather general data representation strategy from optimization of a fundamental physical limit to information processing.
This approach can be pursued in a broader context. 

Thermodynamic efficiency arguments apply, for instance, to the Boltzmann machine \cite{BM}, another staple machine learning algorithm. Input patterns drive this neural network out of its parameter dependent equilibrium state, $q_\theta$, to a non-equilibrium state, $p$. {The associated additional free energy, $F_{\rm add} = k T {\cal D}[p\|q_\theta]$ \cite{shaw1984dripping, GavSchul97, Crooks07, takara2010generalization, deffner2016quantum, lostaglio2015description} (${\cal D}$ denotes the Kullback-Leibler divergence), has to be dissipated during the relaxation process involved in predicting labels on new patterns. Those parameters $\theta$ are found that minimize ${\cal D}[p\|q_\theta]$ \cite{BM}, thereby minimizing a lower bound on the dissipation encountered during prediction.} 

Besides energy efficiency, there are other important physical limits to information processing, including speed, accuracy, and robustness, as well as fundamental trade-offs between them (e.g. \cite{proesmans2017underdamped}). One can ask which information processing strategies emerge when those limits are optimized. {It is also interesting to see how this extends to quantum systems (such as treated in e.g. \cite{kim2011quantum,deffner2013quantum,park2013heat,lorenzo2015landauer,weilenmann2016axiomatic,goold2016role,strasberg2017quantum,deffner2017quantum,abah2019implications}), and to investigate the consequences of considering} single-shot quantities (see e.g. \cite{datta2013one, faist2018fundamental} and references therein). 

\begin{acknowledgments} 
I am most grateful for support from the Foundational Questions Institute (Grants No. FQXi-RFP3-1345, and FQXi-RFP 1820 with the Fetzer Franklin Fund). I thank Rob Shaw for patiently providing feedback and inspiration, and Elan Stopnitzky for commenting on earlier versions of this manuscript. Helpful discussions with John Bechhoefer, Tobi Berger, Bill Bialek, Gavin Crooks, Josh Deutsch, Arne Grimsmo, Massimiliano Esposito, Chris Jarzynski, Matteo Marsili, Marcelo Magnasco, Seth Lloyd, Tom Ouldridge, and David Sivak were greatly appreciated.
\end{acknowledgments} 

\bibliographystyle{unsrt}
\bibliography{TCBM-citations-clean}
\end{document}